\documentclass[10pt]{iopart}
\expandafter\let\csname equation*\endcsname\undefined
\expandafter\let\csname endequation*\endcsname\undefined
\usepackage{amsmath, amssymb}
\usepackage{graphicx}
\usepackage{epstopdf}
\usepackage{units}
\usepackage{cite}
\newcommand{\ie}{\textit{i.e.,}\ }

\newcommand{\md}{\mathrm{d}}
\newcommand{\mcr}{\mathrm{cr}}

\emergencystretch=\maxdimen
\begin{document}

\title{Creating pair plasmas with observable collective effects}

\author{Kenan Qu$^1$, Sebastian Meuren$^2$, Nathaniel J. Fisch$^1$}
\address{$^1$ Department of Astrophysical Sciences, Princeton University,  Princeton, New Jersey 08544, USA }  
\address{$^2$ Stanford PULSE Institute, SLAC National Accelerator Laboratory, Menlo Park, California 94025, USA } 

\vspace{10pt}
\begin{indented}
	\item[] \today
\end{indented}

\begin{abstract}
	Although existing technology cannot yet directly produce fields at the Schwinger level, experimental facilities can already  explore  strong-field QED phenomena by taking advantage of the Lorentz boost of energetic electron beams. Recent studies show that QED cascades can create electron-positron pairs at sufficiently high density to exhibit collective plasma effects. 
	Signatures of the collective pair plasma effects can appear in exquisite detail through plasma-induced frequency upshifts and chirps in the laser spectrum. Maximizing the magnitude of the QED plasma signature demands high pair density and low pair energy, which suits the configuration of colliding an over $\unit[10^{18}]{Jm^{-3}}$ energy-density electron beam with a  $\unit[10^{22}\mathrm{-}10^{23}]{Wcm^{-2}}$ intensity laser pulse. The collision creates pairs that have a large plasma frequency, made even larger as they slow down or reverse direction due to both the radiation reaction and laser pressure.  
	This paper explains at a tutorial level the key properties of the QED cascades and laser frequency upshift, and at the same time  finds the minimum parameters that can be used to produce observable QED plasma. 
\end{abstract}

%
%
%
%
\ioptwocol
\thispagestyle{empty}

\section{Introduction}

According to QED theory, when the field exceeds the Schwinger limit~\cite{Schwinger_1951} $E_\mcr$, the quantum vacuum becomes unstable and it spontaneously creates pairs of electrons and positrons. The oppositely charged electrons and positrons at high density naturally lead to collective plasma effects in the so-called ``QED plasma'' regime~\cite{uzdensky_plasma_2014, grismayer_laser_2016, uzdensky_extreme_2019, zhang_relativistic_2020, Qu_QED2021, Qu_QED2022}. QED plasma effects dominate in astrophysical environments like near a black hole~\cite{RUFFINI20101} or magnetar~\cite{kaspi_magnetars_2017, cerutti_electrodynamics_2017}. Our current understanding of these environments~\cite{ZhangBing2020} is based upon strong-field QED theory for pair creation and plasma theory for the subsequent pair-pair interactions. However, to accurately describe how the QED pair plasmas emit observable radiation and affect the information delivery in the cosmological horizon, it is critical to address how the collective plasma and strong-field QED processes interplay. 

Recent progress in the study of QED physics has been stimulated by the advances of high-power laser technology. Since the invention of chirped-pulse amplification~\cite{cartlidge_light_2018, OPCPA_2019, danson_2019}, the record laser intensity~\cite{Yoon2019} has grown steadily from $\unit[10^{15}]{Wcm^{-2}}$ to $\unit[10^{23}]{Wcm^{-2}}$. Although the latter number is still six orders of magnitude lower than needed for providing $E_\mcr$, we can bridge the gap by colliding the laser with an energetic electron beam. The ultra-relativistic electrons boost the laser field by  orders of magnitude in the electron rest frame, making it possible for existing lasers to test quantum effects. Applying this method, the seminal Stanford E-144 experiment~\cite{bula_observation_1996,burke_positron_1997} in the 1990s detected evidence of positron creation using a $\unit[10^{18}]{Wcm^{-2}}$ laser colliding with a near $\unit[50]{GeV}$ electron beam. The quantum nonlinearity parameter, defined as the ratio of the field to the critical field, is $\chi=E^*/E_\mcr \sim 0.3$ ($E^*$ is measured in the electron rest frame) for this experiment. 
Two decades later, the Gemini laser facility~\cite{cole_experimental_2018, poder_experimental_2018} employed a $\unit[4\times10^{20}]{Wcm^{-2}}$ laser pulse colliding with a GeV electron beam, created via laser wakefield acceleration (LWFA), to observe signatures of quantum radiation reaction at $\chi \sim 0.1$. The commissioned E-320 experiment~\cite{E320} is designed to extend the Stanford experiment and collide a $\unit[10^{20}]{Wcm^{-2}}$ laser with $\unit[10]{GeV}$ electron beam to reach $\chi\sim 1$.

While the community is focusing on testing QED effects at the single particle level, we note that the technology for accessing the QED plasma regime is, in fact, already available~\cite{Qu_QED2021, Qu_QED2022}. Suppose we can colocate a $\unit[10^{23}]{Wcm^{-2}}$ laser with the $\unit[30]{GeV}$ electron beam at SLAC~\cite{FACET-II-TDR, Meuren_2020}, then the $\chi$ parameter  reaches $\sim100$ which is sufficient to produce a QED cascade~\cite{di_piazza_extremely_2012, sokolov_pair_2010, hu_complete_2010, thomas_strong_2012, neitz_stochasticity_2013, bulanov_electromagnetic_2013, blackburn_quantum_2014, green_transverse_2014, vranic_all-optical_2014, blackburn_scaling_2017, vranic_multi-gev_2018, magnusson_laser-particle_2019}. Such a cascade, shown in recent numerical simulations~\cite{Qu_QED2021, Qu_QED2022}, creates pairs at sufficiently high density and low energy that the collective plasma effects begin to show signatures during the laser-pair interaction. 

However, creating a QED plasma  and probing its collective effects, while technically possible, is not so simple. First, the created pairs gain high energy either directly from the gamma photons which they decay from or from the strong laser field. The high pair energy means an increased relativistic mass which significantly suppresses their contribution to collective plasma effects. Second, even with extreme parameters such as a $\unit[10^{23}]{Wcm^{-2}}$ laser and a $\unit[1]{nC}$, $\unit[30]{GeV}$ electron beam, the created pair plasma only has a charge of $\sim \unit[100]{nC}$ distributed in micron scale. The low charge number and small volume prohibit the onset of most plasma instabilities. Third, the pair particles are subject to the ponderomotive force of the intense laser and they undergo rapid volume expansion. Already traveling at relativistic speeds, pair particles last as a plasma within the laser only for picoseconds as numerically demonstrated in~\cite{Qu_QED2021, Qu_QED2022}. 

Thus, detecting the subtle collective effects of QED plasma requires  methods that are sensitive and robust. In views of  the aforementioned challenges, we suggest~\cite{Qu_QED2021, Qu_QED2022}  employing a $\unit[10^{23}]{Wcm^{-2}}$ laser to collide with a \textit{dense} high energy electron beam. The induced QED cascade can not only produce pairs at high density but also low energy. Both properties contribute to strong collective plasma effects. More importantly, the laser pulse, while creating the QED cascade, also probes the time varying pair plasma through the induced frequency change~\cite{morgenthaler1958velocity, wilks1988frequency, esarey_frequency_1990, PRL_Wood_1991, mendoncca2000book, Mendonca2002, Kenan_slow_ionization, Shvets2017,Nishida2012, Kenan_2018_upshift,Edwards_Chirped,Bulanov2005, Peng_PRApp2021}. The laser frequency upshift, determined solely by the change of plasma frequency, provides a robust and unambiguous signature of the collective plasma effects. 

In this paper, we will elaborate on the joint production-observation problem of collective effects of QED plasmas. We analyze the available technologies and assess their advantages for producing high-density and low-energy pair plasma. In Sec.~\ref{Sec:Energy}, we compare the laser-laser collision approach and the beam-laser collision approach for creating plasma and for reducing the relativistic boost of the pair mass. In Sec.~\ref{Sec:Density}, we find the condition on energy density of the electron beam that can create an observable pair plasma. For providing the electron beam, we show the availability of existing conventional electron beam facilities and the promise of the LWFA method at high-power laser facilities. In Sec.~\ref{Sec:Upshift}, we explain in detail how the laser frequency spectrum changes in a time-varying plasma and derive the amount of laser frequency upshift. In Sec.~\ref{Sec:concl}, we present our conclusion.

\section{Reducing the pair energy for strong plasma signatures}\label{Sec:Energy}

The plasma frequency is determined by both the pair density $n_p$ and pair energy (proportional to its Lorentz factor $\gamma$): $\omega_p = \sqrt{n_p e^2/(\epsilon_0^2\gamma m_e)}$, where $e$ is the natural charge, $\epsilon_0$ is the vacuum permittivity, and $m_e$ is the pair rest mass. It is thus key to prepare QED pairs at low energy for detecting their collective effects. Otherwise, high particle energy causes large pair mass from relativistic effects and would substantially suppress their collective response. The requirement of low pair energy seems to conflict with the QED condition that gamma photon emission takes place only with high energy particles. This is true with the laser-laser collision approach for reaching the QED regime, but the conflict is avoided in an electron-beam driven QED cascade.

\subsection{Laser-laser collision cascade}

A laser-laser collision approach of QED cascade, also referred to as the ``avalanche-type'' cascade, employs two ultra-intense counterpropagating laser pulses overlapping in a region with stationary seed electrons~\cite{jirka_electron_2016, grismayer_seeded_2017}. The strong laser beat  wave accelerates the electrons to relativistic velocities. As the electron Lorentz factor $\gamma$ increases, the laser field is boosted by an increasing factor to reach the quantum critical field. Once the quantum nonlinearity parameter $\chi=\gamma E/E_\mcr$ reaches near unit value, the electrons begin to emit high energy gamma photons that can decay into electron-positron pairs. The pairs are then accelerated by the laser field to continue the QED process and develop into a cascade. This process is ``self-sustained'', \ie it terminates only when the pairs escape the laser focal region. 

To reach the QED cascade condition, the laser-laser collision approach~\cite{grismayer_laser_2016, grismayer_seeded_2017} likely requires $\unit[10^{24}]{Wcm^{-2}}$ laser intensities, corresponding to laser amplitude $a_0\equiv eE/(m_ec^2\omega_0) \sim10^3$,  where $\omega_0$ is the laser frequency. If a pair plasma is created, the pair particles would be quickly accelerated to high energy with Lorentz factors $\gamma>10^3$. Thus, their contribution to the plasma frequency would be suppressed by a factor of at least $10^3$. The smallness of their contribution means that detecting the collective plasma effects would need  higher pair density which in turn requires even stronger lasers. Moreover, because of the high pair energy, the contribution of the pairs to the collective plasma effects could be less than that of the stationary seeding electrons unless the pair number multiplication factor is larger than $\gamma>10^3$.

\subsection{Electron-beam driven cascade}
In contrast to the laser-laser collision approach, the electrons in a beam-driven QED cascade begin with the maximum particle energy. Once the ramping-up laser intensity reaches $\chi=2\gamma E/E_\mcr\geq1$ (the factor of $2$ arises from the counterpropagating configuration), the electrons begin to emit gamma photons and lose significant energy. Electron-positron pairs are created by acquiring the energy of the emitted gamma photons. If the pairs have sufficiently high Lorentz factors, \ie $\chi \geq 1$, they emit more gamma photons that can decay into more pairs. This process is thus also called the ``shower-type'' QED cascade. 
This type of cascade converts electron beam energy into pair particles during its collision with a strong laser. The laser pulse, however, does not contribute to the pair energy. The created pairs exhibit increasingly strong plasma behavior both when their density grows and when their energy decreases. 
This approach takes advantage of the high beam energy available through existing electron beam facilities; hence, it greatly reduces the required laser intensity. For example with $\unit[30]{GeV}$ electron beam energy, $\unit[10^{20}]{Wcm^{-2}}$ laser intensity could already reach $\chi\geq 1$ and produce pair number multiplication. Higher laser intensity at $\unit[10^{22} \mathrm{-}10^{23}]{Wcm^{-2}}$, combined with the same electron beam, could reach the extreme quantum limit $\chi\gg 1$ and induce a full-featured QED cascade~\cite{Qu_QED2021, Qu_QED2022}. 

The low requirement for laser intensity not only avoids the technical challenges of building $\unit[100]{PW}$-class laser, but also allows the pairs to exhibit strong plasma effects. In the electron-beam driven cascade, the counterpropagating laser pulse decelerates the particles to reduce the pair energy. This means that the relativistic particle mass decreases and their contribution to the plasma frequency increases. The minimum pair energy (and hence the maximum contribution to plasma frequency) is achieved if the pairs could be fully stopped, at least, in the longitudinal direction. In the ``pair-stopping'' regime, the minimum pair energy is then determined solely by their transverse quiver motion driven by laser, and thus $\gamma\sim a_0$ for $a_0\gg 1$. 

Reaching the ``pair-stopping'' regime requires the laser amplitude to exceed the threshold value: $a_{0,\mathrm{th}} {\approx} 100$ corresponding to $I_{0,\mathrm{th}}  \approx \unit[10^{22}\mathrm{-}10^{23}]{Wcm^{-2}}$ for $\mu m$-wavelength lasers. The threshold laser amplitude is obtained~\cite{Qu_QED2021, Qu_QED2022} by analyzing the two dominating mechanisms of pair deceleration. The high energy pairs first lose energy mainly through the quantum radiation reaction which terminates when the pair energy decreases below the value for $\chi (\propto a_0\gamma)\lesssim 0.1$. Then the second mechanism---the ponderomotive force of the counterpropagating laser---begins to dominate the pair deceleration. The ponderomotive pressure can reduce the  longitudinal electron momentum by the maximum amount of $\gamma\cong a_0$ in the limit of a single laser wavelength~\cite{di_piazza_extremely_2012}, and this value is slightly larger for longer laser pulses~\cite{Griffith2022}. 
These two mechanisms scale with $a_0$ differently. 
By equating the terminal pair energy for quantum radiation reaction and the maximum pair energy that can be exchanged with the laser field, we can find the threshold laser amplitude: $a_{0,\mathrm{th}}\approx 100$. Above the threshold, the pair particles could be fully stopped  reaching the minimum longitudinal momentum.

If the laser intensity substantially exceeds $I_{0,\mathrm{th}}$, some of the pair particles, if they remain near the laser center, could be reaccelerated by the strong ponderomotive force towards the laser beam direction. The reacceleration on one hand side increases the pair Lorentz factor, but on the other hand side also reduces the laser frequency in the copropagating pair rest frame. For the particular plasma signature of laser frequency upshift, it is shown~\cite{Griffith2022} that reacceleration can accentuate the amount of frequency upshift by up to a factor of $2$.

\section{Reaching high pair density for large plasma effects}\label{Sec:Density}

In an electron-beam driven QED cascade, all the pairs are created by converting the energy of either the electron beam or the pairs created by it, mediated by high energy gamma photons. Since the energy contribution from the laser  and long-wavelength emissions are both negligible, the total particle energy is conserved during the cascade. In other words, the integrated particle energy-density over the whole space is conserved.

The conservation of integrated particle energy-density means that creating high density pair plasma requires employing a high energy-density electron beam. Quantitatively, the final pair density $n_p$ can be estimated as   
\begin{equation}
	n_p\approx n_0\chi_0,
\end{equation}
where $n_0$ is the density of injected electrons and $\chi_0\approx 2a_0 \gamma_0 (\hbar\omega_0)/(m_ec^2)$, interpreted as the pair multiplication factor, is the quantum nonlinearity parameter for the injected electron beam with $\gamma_0$ in the laser field. This relation assumes that all the pair particles interact with constant laser intensity and the cascade terminates at $\chi\sim1$ when their emitted photons can no longer decay into more pairs. For  $\mu m$-wavelength lasers, the relation can be written numerically as $n_p \approx 4\times10^{-6} a_0\gamma_0 n_0$. 
Thus, for the cascade to create a pair density near the critical density $n_p\sim\unit[2\times10^{21}]{cm^{-3}}$, the electron beam needs to have energy density $\gamma_0n_0 \sim \unit[10^{25}]{cm^{-3}}$ assuming that the laser reaches at the ``pair-stopping'' threshold amplitude ($a_0\approx 100$). 

Note that, although employing a higher laser intensity can improve the pair multiplication factor, it does not increase the pair plasma frequency. Once the laser amplitude is  above $a_{0,\mathrm{th}}$, the final pair motion becomes dominantly transverse with kinetic energy proportional to the laser amplitude. Higher laser amplitude simultaneously induces a larger pair multiplication factor and a larger Lorentz factor, canceling their contribution to the plasma frequency $\omega_p\propto \sqrt{n_p/\gamma}$.


The required high energy-density $\gamma_0n_0{\sim} \unit[10^{25}]{cm^{-3}}$ naturally favors conventional accelerators for their high luminosity. For GeV-level electron beam energy, the density needs to reach $\unit[10^{19}]{cm^{-3}}$. For example, the nC-level electron charge is accessible in several electron accelerator facilities including SLAC, eRHIC, ILC, CLIC, etc. Taking into account the beam bunch size, their electron densities all exceed $\unit[10^{19}]{cm^{-3}}$. Notably, their beam energy are in the range of $\unit[10]{GeV}$ to TeV level enabling $\gamma_0n_0 \sim  \unit[10^{27}-10^{30}]{cm^{-3}}$.

Laser wakefield acceleration is an alternative technique which  yields hundreds-of-MeV to GeV-level electron beams at high-power laser facilities. It uses the ponderomotive force of a strong laser pulse to push electrons in a plasma medium via either  self-modulated beat wave or a hollow bubble. 
Present LWFA techniques, however, have the major drawback of a trade-off between high beam energy or high charge number. The current record~\cite{PRL_LWFA2019} for LWFA electron energy is $\sim\unit[8]{GeV}$, but it only has $\sim\unit[5]{pC}$ total charge. The energy density of this electron beam is still three orders of magnitude lower than the required value.  Higher charge number could be achieved only by compromising the beam length and more importantly the beam energy, which both reduce the energy density. Recent studies~\cite{Andreev2003, Welch2017, Kumar2021, Brunetti2022} show via numerical simulation that long-wavelength CO$_2$ lasers at high power might overcome the energy-density barrier and produce high electron charge number at the GeV level through LWFA. Nevertheless, producing $\unit[1]{nC}$ of electrons at $\unit[10]{GeV}$, which contains $\unit[10]{J}$ electron kinetic energy, will need next generation laser technology capable of delivering $\unit[100]{J}-\unit[1000]{J}$ laser pulses even at $1\%-10\%$ energy conversion efficiency.


\section{Laser frequency upshift induced by plasma effects}\label{Sec:Upshift}

If a pair plasma were created  through the QED cascade as we described above, it would be micrometer sized with relativistic velocity making diagnosing it challenging. Detecting the subtle collective effects needs unconventional methods that are sensitive and robust. One of the lowest order plasma effects is the dispersion relation. As the pair plasma is formed, the plasma frequency grows both when the pair density increases and when the pair energy decreases. 
The growing plasma frequency changes the dispersion relation of laser by reducing the refractive index and increasing the laser phase velocity. Sudden creation of plasma over space amounts to a temporal interface of refractive indices, through which the laser frequency is upshifted. 
Considering that the pair plasma dimension is only a fraction of the laser duration, the increased laser phase velocity also causes its wavefront to compress towards the front which can be detected as a chirp in the laser spectrum. Both the laser frequency upshift and chirp arise from the temporal evolution of the plasma frequency, hence they serve as unambiguous signatures of collective effects. 

The creation of pair plasma is modeled as a temporal interface of refractive indices, which is known to cause laser frequency upshift~\cite{felsen1970wave, kravtsov1974geometrical, AuYeung:83, stepanov1993waves, Mendonca2002}.  The frequency upshift process~\cite{wilks1988frequency, esarey_frequency_1990, PRL_Wood_1991, Kenan_slow_ionization, Shvets2017,Nishida2012, Kenan_2018_upshift,Edwards_Chirped,Bulanov2005, Peng_PRApp2021} is analogous to the trivial process of laser wavelength shift when crossing a spatial interface of refractive index. The concept of laser frequency change in dynamic media was first studied~\cite{morgenthaler1958velocity} by Morgenthaler in 1958. With rapidly growing laser technology in the 1970s, it is found~\cite{ostrovskii1971review, jiang1975wave, lampe1978interaction} that laser-breakdown plasmas can serve as such dynamic media. The concept was further developed as, so-called, ``photon accelerators''~\cite{wilks1989photon, mori1991generation, kalluri1992frequency, savage1993frequency, mendoncca2000book}, in which the laser propagates in the rear edge of a plasma wave wakefield. Since the laser co-moves in a positive density gradient, it can be frequency upconverted continuously. Using laser-induced ionization, frequency upconversion has been experimentally demonstrated in the microwave~\cite{yablonovitch1973spectral, joshi1990demonstration, kuo1990frequency, savage1993frequency, yugami2002experimental}, terahertz~\cite{Nishida2012} and optical~\cite{Suckewer2002, Suckewer2005} regimes. 

In a QED cascade, the created pair plasma interacts with the laser in a manner similar to an ionization front but in a counterpropagating geometry. It changes the refractive index in both space and time, and leads to changes in both laser frequency and wavelength. In the following, we will first pictorially explain the change of laser spectrum  using a spacetime diagram and then analytically derive the amount of upshift due to the transient and inhomogeneous pair plasma. 

\subsection{Diagram explanation of laser frequency upshift}

\begin{figure}[ht]
	\centering
	\includegraphics[width=\linewidth]{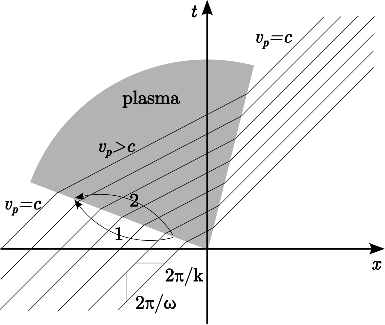}
	\caption{Spacetime diagram of plasma creation and laser frequency upshift. The phase differences are identical for both path 1 and path 2.} 
	\label{diag}
\end{figure}

Laser propagation  can be illustrated using the spacetime diagram, as shown in \fref{diag}. The shaded area in \fref{diag} represents pair plasma which grows in time and expands in space. The parallel lines represent the laser wavefront propagating in the $x$-direction. The vertical and horizontal spacings of the lines correspond to the laser frequency and wavevector, respectively. 
As the laser propagates through the vacuum-plasma interface, its phase velocity changes from $c$ to $v_p= c/\sqrt{1-\omega_p^2 /\omega^2}>c$. 
The phase of the laser is nevertheless continuous across the interface, represented as non-broken lines in \fref{diag}. 

The change of laser frequency and wavenumber results from both the change of phase velocity, denoted as the slope change of the parallel lines in \fref{diag}, and the angle of interface. The interface can be categorized in the following types depending on its angle: 
\begin{enumerate}
	\item A spatial interface of media is represented by a vertical boundary parallel to the $t$-axis in  spacetime diagram. The laser wavefront when crossing the spatial interface conserves its vertical spacing, \ie its frequency; its horizontal spacing changes correspondingly, indicating a change in wavenumber. 
	\item A temporal interface of media is represented by a horizontal boundary parallel to the $x$-axis in spacetime diagram. The laser wavefront when crossing it conserves its horizontal spacing but changes its vertical spacing, indicating a change in frequency. 
	\item More generally, if the interface involves both spatial and temporal changes of refractive index, it is represented in the spacetime diagram by a boundary that is not parallel to either $t$- or $x$-axis, the laser wavefront spacing changes in both directions, indicating changes in both frequency and wavevector. 	
\end{enumerate}
Because the laser phase is continuous, any separation on the interface has identical optical paths in both media, leading to the identity
\begin{equation}
	k_1\Delta x - \omega_1\Delta t = k_2\Delta x - \omega_2\Delta t,
\end{equation}
where $\omega_i$ and $k_i$ are the frequency and wavevector in the $i$'th medium, and $\Delta t$ and $\Delta x$ are arbitrary spacetime distances on the interface. The slope of interface is most conveniently described by the parameter $1/\beta= c\Delta t/\Delta x$. The parameter $\beta c$ can also be interpreted as the velocity of the interface. Then using the relation $v_{p,i}=\omega_i/k_i$, we can obtain
\begin{equation} \label{freqchange1}
	\begin{aligned}
	\omega_2 &= \left(\frac{\beta^{-1} - c/v_{p1}}{\beta^{-1} - c/v_{p2}} \right) \omega_1, \\
	k_2 &= \left(\frac{v_{p1}/c-\beta}{v_{p2}/c-\beta} \right) k_1.
	\end{aligned}
\end{equation}
These relations describe how the frequency and wavevector change when the laser propagates through a spacetime interface moving at velocity $v=c\beta$. The shifts of frequency and wavevector can then be expressed as
\begin{equation} \label{freqchange2}
	\begin{aligned}
	\Delta\omega &= \left(\frac{ c/v_{p2} - c/v_{p1}}{\beta^{-1} - c/v_{p2}} \right) \omega_1, \\
	\Delta k &= \left(\frac{v_{p1}/c-v_{p2}/c}{v_{p2}/c-\beta} \right) k_1.
	\end{aligned}
\end{equation}
The process of interface crossing can take place either when the laser propagates faster than the interface ($v_{p1,2}>\beta c$) or when the interface overtakes the laser ($\beta c>v_{p1,2}$). But the parameter regime $v_{pi}>\beta c>v_{pj}$ ($i\neq j$) forbids laser propagation after it crosses the interface, and hence is nonphysical. 

\begin{figure}[ht]
	\centering
	\includegraphics[width=\linewidth]{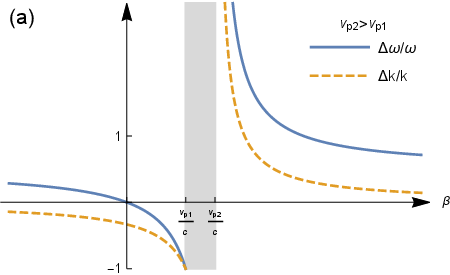}\\ \vspace{6pt}
	\includegraphics[width=\linewidth]{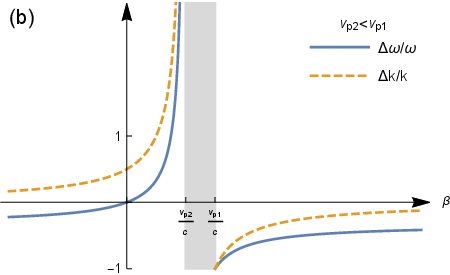}
	\caption{Frequency change ($\Delta\omega$, solid blue curves) and wavevector change ($\Delta k$, dashed orange curves) for different interface velocities $\beta$ assuming (a) $v_{p2}>v_{p1}$ and (b) $v_{p2}<v_{p1}$. The shaded region is nonphysical because the laser cannot propagate in the second medium after crossing the interface. } 
	\label{upshift}
\end{figure}

The amount of frequency shift ($\Delta\omega$) and wavevector shift ($\Delta k$) with varying interface velocity $\beta$ is plotted in \fref{upshift} assuming, respectively, (a) $v_{p2}>v_{p1}$ and (b) $v_{p2}<v_{p1}$. Depending on the relations of the interface and laser velocities, the plot can be divided into four regimes, among which the shaded areas are nonphysical. 

A subluminal copropagating interface $v_{p1,2}>\beta c\geq0$ traverses through the laser pulse from the laser front to laser tail. If $v_{p2}>v_{p1}$, the laser wavefront propagates faster after crossing the interface and it leads to an increase of wavelength and period. Thus, both the laser frequency and wavevector are downshifted. 
In the limit of $\beta \to 0$, it reduces to a stationary interface which downshifts the laser wavevector by $v_{p1}/v_{p2}$ but  does not change the laser frequency. 
As the interface velocity increases, the slower relative motion between the laser wavefront and the interface lengthens the wavefront spreading process, thereby amplifying the downshifts.

A superluminal copropagating interface $\beta c> v_{p1,2} >0$ traverses through the laser pulse from the laser tail to front. For $v_{p2}>v_{p1}$, the faster phase velocity in the tail compresses the laser wavefront. It leads to a decrease of wavelength and period, and hence an upshift of laser frequency and wavevector. Similarly to a subluminal interface, a smaller relative interface-to-laser velocity lengthens the time of wavefront compression. Thus, the frequency and wavevector upshifts become greater as $\beta c\to v_{p2}$. In the case of a laser crossing a sudden and homogeneous interface $\beta\to\infty$, the spatial separation of the laser wavefront, or wavelength $\lambda$, does not change, \ie $\Delta k=0$, but the temporal separation is reduced from $\lambda/v_{p1}$ to $\lambda/v_{p2}$ so the frequency is upshifted by a factor $v_{p2}/v_{p1}$.

A counterpropagating interface $\beta<0$ traverses through the laser pulse from the laser front to tail. Similar to the scenario of a subluminal copropagating interface, the laser wavefront, which has a faster phase velocity in the front, is lengthened. This causes a downshift of wavevector, $\Delta k<0$. From the time point of view, the laser wavefronts in the counterpropagating configuration cross the interface at a rate higher than the laser frequency. This allows the laser tail to propagate more time at $v_{p2}$ ($>v_{p1}$) than the front for the same distance, similar to the effect of a superluminal copropagating interface. Thus, the laser wavefront is compressed in time and the laser frequency is upshifted.

In an electron-beam driven QED cascade, the laser pulse crosses the vacuum-plasma interface twice, when entering and exiting the plasma. The first encounter occurs when the laser pulse and electron beam begin to collide. The pairs are initially created inside the electron beam and thus the vacuum plasma interface has the same Lorentz factor with the beam, \ie $\beta_1 \approx -1$. (The $\beta$ factor could locally exceed unit value considering the fact that the pair density spacetime gradient is determined by both the particle density and laser intensity. But the asymptotic speed of the pair plasma front equals to that of the electron beam.)  If we assume a homogeneous plasma, the laser phase velocity changes from $c$ to $v_p=c\Big/\sqrt{1-\omega_p^2/\omega^2}>c$ after crossing the interface. According to \eref{freqchange1}, the laser frequency and wavevector change to $\omega_2=2\omega/(1+c/v_p)$ and $k_2=2k/(1+v_p/c)$, respectively. The created pairs lose most of their energy and are subject to the ponderomotive potential of the strong laser pulse. As explained in the last section, the pairs are mostly stopped and partially reflected while expanding in transverse directions. Also, because the fast moving pairs have high energy and hence contribute little to the plasma frequency, we can describe the second plasma-vacuum interface with $\beta\sim0$. Thus, the laser frequency does not change and the wavevector changes as $k_f=k_2(v_p/c).$ Therefore, the vacuum-plasma-vacuum interfaces change the laser frequency and wavevector as
\begin{equation}\label{diag-upshift}
	\begin{aligned}
	\omega_f &= \left(\frac{2}{1+c/v_p}\right) \omega  \approx \omega + \frac{\omega_p^2}{4\omega}  , \\
	 k_f &= \left(\frac{2}{1+c/v_p}\right) k \approx k + \frac{c\omega_p^2}{4\omega}  .
	\end{aligned}
\end{equation}
Equations~\eref{diag-upshift} show that the amount of laser frequency upshift is $\omega_p^2/(4\omega)$. It is lower than the laser frequency upshift in sudden ``flash'' ionization by a factor of $2$ caused by the finite velocity of the interface. 
The laser frequency change could be measurable if the pair plasma density needs to reach a non-negligible fraction of the laser frequency. Assuming laser amplitude $a_0\sim 100$, the pair density needs to reach $\unit[10^{21}]{cm^{-3}}$.

\subsection{Chirp of laser spectrum caused by QED cascade}

\begin{figure}[ht]
	\centering
	\includegraphics[width=\linewidth]{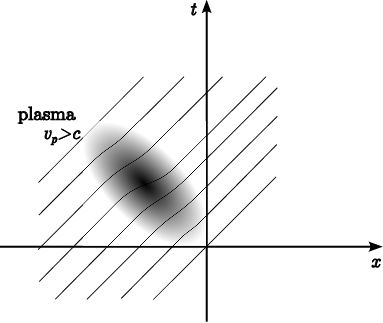}
	\caption{Spacetime diagram of plasma creation and laser frequency upshift. } 
	\label{diag-qed}
\end{figure}

The above analysis assumes homogeneous plasma frequency to obtain equation~\eref{diag-upshift}. However, the combined processes of pair creation and volume expansion cause the plasma density to be inhomogeneous in both space and time. We illustrate the interaction of the laser and pair plasma in \fref{diag-qed}. The diagram shows that as the laser pulse enters and exits the plasma-vacuum interfaces,  each part of the laser pulse propagates through plasma at different velocities. Since only the laser center propagates through the  densest part of plasma, it experiences the largest frequency and wavevector upshifts. Therefore, the laser pulse is chirped.

The chirp profile can be found by tracing the amount of phase shift when the laser propagates through the inhomogeneous plasma. Since the phase shift is different for each part of the laser pulse, it is convenient to define $\xi=x-ct$ denoting the relative delay from the laser front and $\tau=t$ denoting the propagation time. The laser phase can then be written as $\phi=\omega(t-x/v_p) =- \omega\xi/v_p+ \omega(1-c/v_p)\tau$. 
The expression in the $(\xi,\tau)$ coordinate separates the laser phase into its internal phase variation and the induced changes along $\tau$. For laser propagating in vacuum, $\phi=-\omega\xi/c$ which is a constant along $\tau$. If the laser propagates through plasma as shown in \fref{diag-qed}, the collective plasma effect causes a phase shift $\md\phi= (1-c/v_p) \md\tau$, which accumulates in $\tau$. 
For small plasma frequencies, $1-c/v_p \approx \omega_p^2/(2\omega)$. Each part of laser at $\xi$ propagates through plasma at $(\xi+c\tau',\tau')$ over the range $-\infty <\tau'<\tau$. Thus, the total phase shift can be found as
\begin{equation}
	\Delta\phi= \int_{-\infty}^\tau \omega_p^2(\xi+\tau',\tau')/(2\omega)\md \tau'. 
\end{equation}
Neglecting the small change of $1/\omega$ and transforming back to the $(x,t)$ coordinate, we have 
\begin{equation}
	\Delta\phi= \frac{1}{2\omega}\int_{-\infty}^t \omega_p^2(x-ct+ct',t')\md t'. 
\end{equation}
The frequency and wavevector after propagating through inhomogeneous plasma can thus be expressed as
\begin{align}
	\Delta\omega(x,t) &= \frac{\partial\Delta\phi}{\partial t} \nonumber \\
	&\!\!\!\!\!\!\!\!\!= \frac{\omega_p^2(x,t)}{2\omega} + \frac{1}{2\omega} \int_{-\infty}^t [\partial_t \omega_p^2(x-ct+ct',t')]\md t',  \\
	\Delta k(x,t) &= -\frac{\partial\Delta\phi}{\partial x}  \nonumber \\
	&\!\!\!\!\!\!\!\!\!=  -\frac{1}{2\omega} \int_{-\infty}^t [\partial_x \omega_p^2(x-ct+ct',t')]\md t'.
\end{align}
Note that $\partial_t \omega_p^2= -c\partial_x \omega_p^2$, so the dispersion relation $\Delta\omega - c\Delta k= \omega_p^2/(2\omega)$ is automatically verified. 
We can further simplify the expressions by noting that $\displaystyle \omega_p^2(x,t) = \int_{-\infty}^t [\partial_{t'} \omega_p^2(x-ct+ct',t')]\md t'$ and $(\partial_t + \partial_{t'}) \omega_p^2(x-ct+ct',t') = [\partial_T \omega_p^2(X,T)]^{T=t'}_{X=x-ct+ct'}$, then we obtain the expressions reported in~\cite{Qu_QED2021,Qu_QED2022}
\begin{align}
	\Delta\omega(x,t) &= \frac{1}{2\omega} \int_{-\infty}^t [\partial_T \omega_p^2(X,T)]^{T=t'}_{X=x-ct+ct'}\md t',  \\
	\Delta k(x,t) &=  -\frac{1}{2\omega} \int_{-\infty}^t [\partial_X \omega_p^2(X,T)]^{T=t'}_{X=x-ct+ct'}\md t'.
\end{align}

The expressions show a very intuitive picture: the change of laser frequency and wavevector are caused by the integration of temporal and spatial change of plasma frequency calculated at the retarded position $X=x-c(t-t')$. If the plasma moves with velocity $-c$, then $\partial_T \omega_p^2(X,T) = -c\partial_X \omega_p^2(X,T)$ and hence the amounts of  frequency and wavevector upshift only differ by a factor of $c$. 

Because the laser chirp is related to the rapidness of the plasma frequency change $\partial_T \omega_p^2(X,T)$, the signal could be much larger than the laser frequency shift for small plasma size. In the aforementioned QED cascade, the laser pulse has a typical duration of $\unit[100]{fs}$ corresponding to $\unit[30]{\mu m}$ in length, but the plasma only has $<\unit[10]{\mu m}$ length. Thus, the instantaneous laser frequency upshift could be several times higher than the central frequency change of the whole pulse. In other words, the pair plasma is created when the small electron beam encounters the most intense region of the laser pulse and hence only induces laser frequency upshift near its intensity peak. When averaging over the whole laser pulse, the frequency upshift would decrease by a large factor.

\section{Conclusion}\label{Sec:concl}

The QED plasma dynamics are distinguished from traditional electron-ion plasmas by a number of physical aspects, including special relativistic effects, radiation-reaction effects, and high mobility under laser pressure. Exploiting the  laser frequency upshift relaxes the conditions for QED plasma detection. Thus, creating an observable pair plasma through strong-field QED cascades in terrestrial laboratories becomes possible with state-of-the-art technologies. 

Adopting the electron-beam-laser collision approach, the minimum parameters for testing QED plasma phenomena include laser intensity of $\unit[10^{23}]{Wcm^{-2}}$ and electron beam  energy density of $\unit[10^{18}]{Jm^{-3}}$ ($\gamma n_0\sim\unit[10^{25}]{cm^{-3}}$). 
The required energy density can be readily produced by a conventional electron beam accelerator. Its production at a strong laser facility might also become possible if the LWFA technique can overcome the trade-off between high beam energy and high total electron charge. 
If the high energy-density electron beam is colocated with a PW-class laser, the collision creates QED pairs with growing a density and decreasing energy. In contrast to the direct all-optical laser-laser collision approach, the electron-beam driven QED cascade converts high energy beam  into pairs with low energy and high density, both of which contribute to higher plasma frequency. The use of a high energy electron beam reduces the required laser intensity. The lower laser intensity means that the produced pairs are less energetic, making the plasma frequency larger for the same pair density. 

Identifying the conditions for creating observable QED plasma is timely in view of the present planning of QED  facilities. With current technology, the highest quantum nonlinearity parameter $\chi$  is achieved using conventional electron accelerators. The undergoing Stanford E-320 experiment~\cite{E320} uses a $\unit[10]{GeV}$ beam and a $\unit[10^{20}]{Wcm^{-2}}$ laser to achieve $\chi\sim1$. The electron beam energy density, assuming that the $\unit[2]{nC}$ beam can be compressed to $\unit[0.5]{\mu m} \times (\unit[3]{\mu m})^2$ size, could reach $\gamma_0n_0\sim \unit[10^{25}]{cm^{-3}}$. Creating an observable QED plasma requires an upgrade of the laser by two order of magnitude, reaching $\chi\sim100$. The LUXE experiment at DESY proposes~\cite{DEXY} using a $\unit[17.5]{GeV}$ beam and  $\unit[10^{20}\mathrm{-}10^{21}]{Wcm^{-2}}$ to achieve $\chi\sim1\mathrm{-}3$. The beam at the highest energy configuration is limited to $0.25$nC charge and $\sim50$fs length, hence it needs significant focusing to exhibit collective plasma effects.

\section*{Acknowledgments}
	This work was supported by DOE Grants DE-NA0003871, NNSA DE-SC0021248, and DE-AC02-76SF00515. 

\section*{References}
\bibliography{Upshift_PPCF}

\providecommand{\newblock}{}
\begin{thebibliography}{10}
\expandafter\ifx\csname url\endcsname\relax
  \def\url#1{{\tt #1}}\fi
\expandafter\ifx\csname urlprefix\endcsname\relax\def\urlprefix{URL }\fi
\providecommand{\eprint}[2][]{\url{#2}}

\bibitem{Schwinger_1951}
Schwinger J 1951 {\em Phys. Rev.\/} {\bf 82}(5) 664

\bibitem{uzdensky_plasma_2014}
Uzdensky D~A and Rightley S 2014 {\em Rep. Prog. Phys\/} {\bf 77} 036902

\bibitem{grismayer_laser_2016}
Grismayer T, Vranic M, Martins J~L, Fonseca R~A and Silva L~O 2016 {\em Phys.
  Plasmas\/} {\bf 23} 056706

\bibitem{uzdensky_extreme_2019}
Uzdensky D, Begelman M, Beloborodov A, Blandford R, Boldyrev S, Cerutti B,
  Fiuza F, Giannios D, Grismayer T, Kunz M, Loureiro N, Lyutikov M, Medvedev M,
  Petropoulou M, Philippov A, Quataert E, Schekochihin A, Schoeffler K, Silva
  L, Sironi L, Spitkovsky A, Werner G, Zhdankin V, Zrake J and Zweibel E 2019

\bibitem{zhang_relativistic_2020}
Zhang P, Bulanov S~S, Seipt D, Arefiev A~V and Thomas A~G~R 2020 {\em Phys.
  Plasmas\/} {\bf 27} 050601

\bibitem{Qu_QED2021}
Qu K, Meuren S and Fisch N~J 2021 {\em Phys. Rev. Lett.\/} {\bf 127}(9) 095001

\bibitem{Qu_QED2022}
Qu K, Meuren S and Fisch N~J 2022 {\em Physics of Plasmas\/} {\bf 29} 042117

\bibitem{RUFFINI20101}
Ruffini R, Vereshchagin G and Xue S~S 2010 {\em Physics Reports\/} {\bf 487}
  1--140 ISSN 0370-1573

\bibitem{kaspi_magnetars_2017}
Kaspi V~M and Beloborodov A~M 2017 {\em Annu. Rev. Astron. Astrophys.\/} {\bf
  55} 261

\bibitem{cerutti_electrodynamics_2017}
Cerutti B and Beloborodov A~M 2017 {\em Space Sci. Rev.\/} {\bf 207} 111

\bibitem{ZhangBing2020}
Zhang B 2020 {\em Nature\/} {\bf 587} 45--53

\bibitem{cartlidge_light_2018}
Cartlidge E 2018 {\em Science\/} {\bf 359} 382

\bibitem{OPCPA_2019}
Bromage J, Bahk S~W, Begishev I~A, Dorrer C, Guardalben M~J, Hoffman B~N,
  Oliver J~B, Roides R~G, Schiesser E~M, Shoup~III M~J and et~al 2019 {\em High
  Power Laser Sci\/} {\bf 7} e4

\bibitem{danson_2019}
Danson C~N, Haefner C, Bromage J, Butcher T, Chanteloup J~C~F, Chowdhury E~A,
  Galvanauskas A, Gizzi L~A, Hein J, Hillier D~I and et~al 2019 {\em High Power
  Laser Sci. Eng.\/} {\bf 7} e54

\bibitem{Yoon2019}
Yoon J~W, Jeon C, Shin J, Lee S~K, Lee H~W, Choi I~W, Kim H~T, Sung J~H and Nam
  C~H 2019 {\em Opt. Express\/} {\bf 27} 20412--20420

\bibitem{bula_observation_1996}
Bula C, McDonald K~T, Prebys E~J, Bamber C, Boege S, Kotseroglou T, Melissinos
  A~C, Meyerhofer D~D, Ragg W, Burke D~L, Field R~C, Horton-Smith G, Odian A~C,
  Spencer J~E, Walz D {\em et~al.\/} 1996 {\em Phys. Rev. Lett.\/} {\bf 76}
  3116

\bibitem{burke_positron_1997}
Burke D~L, Field R~C, Horton-Smith G, Spencer J~E, Walz D, Berridge S~C, Bugg
  W~M, Shmakov K, Weidemann A~W, Bula C, McDonald K~T, Prebys E~J, Bamber C,
  Boege S~J, Koffas T {\em et~al.\/} 1997 {\em Phys. Rev. Lett.\/} {\bf 79}
  1626

\bibitem{cole_experimental_2018}
Cole J~M, Behm K~T, Gerstmayr E {\em et~al.\/} 2018 {\em Phys. Rev. X\/} {\bf
  8} 011020

\bibitem{poder_experimental_2018}
Poder K, Tamburini M, Sarri G, Di~Piazza A, Kuschel S, Baird C~D, Behm K,
  Bohlen S, Cole J~M, Corvan D~J, Duff M, Gerstmayr E, Keitel C~H, Krushelnick
  K, Mangles S~P~D {\em et~al.\/} 2018 {\em Phys. Rev. X\/} {\bf 8} 031004

\bibitem{E320}
{Meuren} S 2022 {Light-matter interactions at extreme intensities and
  densities: reaching the Schwinger limit} {\em APS April Meeting Abstracts\/}
  ({\em APS Meeting Abstracts\/} vol 2022) p H07.003

\bibitem{FACET-II-TDR}
{SLAC Site Office} 2016 {\em SLAC-R-1072\/}

\bibitem{Meuren_2020}
Meuren S, Bucksbaum P~H, Fisch N~J, Fiúza F, Glenzer S, Hogan M~J, Qu K, Reis
  D~A, White G and Yakimenko V 2020 On seminal hedp research opportunities
  enabled by colocating multi-petawatt laser with high-density electron beams

\bibitem{di_piazza_extremely_2012}
Di~Piazza A, M\"uller C, Hatsagortsyan K~Z and Keitel C~H 2012 {\em Rev. Mod.
  Phys.\/} {\bf 84} 1177

\bibitem{sokolov_pair_2010}
Sokolov I~V, Naumova N~M, Nees J~A and Mourou G~A 2010 {\em Phys. Rev. Lett.\/}
  {\bf 105} 195005

\bibitem{hu_complete_2010}
Hu H, M\"{u}ller C and Keitel C~H 2010 {\em Phys. Rev. Lett.\/} {\bf 105}
  080401

\bibitem{thomas_strong_2012}
Thomas A~G~R, Ridgers C~P, Bulanov S~S, Griffin B~J and Mangles S~P~D 2012 {\em
  Phys. Rev. X\/} {\bf 2} 041004

\bibitem{neitz_stochasticity_2013}
Neitz N and Di~Piazza A 2013 {\em Phys. Rev. Lett.\/} {\bf 111} 054802

\bibitem{bulanov_electromagnetic_2013}
Bulanov S~S, Schroeder C~B, Esarey E and Leemans W~P 2013 {\em Phys. Rev. A\/}
  {\bf 87} 062110

\bibitem{blackburn_quantum_2014}
Blackburn T~G, Ridgers C~P, Kirk J~G and Bell A~R 2014 {\em Phys. Rev. Lett.\/}
  {\bf 112} 015001

\bibitem{green_transverse_2014}
Green D~G and Harvey C~N 2014 {\em Phys. Rev. Lett.\/} {\bf 112} 164801

\bibitem{vranic_all-optical_2014}
Vranic M, Martins J~L, Vieira J, Fonseca R~A and Silva L~O 2014 {\em Phys. Rev.
  Lett.\/} {\bf 113} 134801

\bibitem{blackburn_scaling_2017}
Blackburn T~G, Ilderton A, Murphy C~D and Marklund M 2017 {\em Phys. Rev. A\/}
  {\bf 96} 022128

\bibitem{vranic_multi-gev_2018}
Vranic M, Klimo O, Korn G and Weber S 2018 {\em Sci. Rep.\/} {\bf 8} 4702 ISSN
  2045-2322

\bibitem{magnusson_laser-particle_2019}
Magnusson J, Gonoskov A, Marklund M, Esirkepov T~Z, Koga J~K, Kondo K, Kando M,
  Bulanov S~V, Korn G and Bulanov S~S 2019 {\em Phys. Rev. Lett.\/} {\bf
  122}(25) 254801

\bibitem{morgenthaler1958velocity}
Morgenthaler F~R 1958 {\em {IEEE} Trans. Microw. Theory Tech.\/} {\bf 6} 167

\bibitem{wilks1988frequency}
Wilks S~C, Dawson J~M and Mori W~B 1988 {\em Phys. Rev. Lett.\/} {\bf 61} 337

\bibitem{esarey_frequency_1990}
Esarey E, Ting A and Sprangle P 1990 {\em Phys. Rev. A\/} {\bf 42} 3526

\bibitem{PRL_Wood_1991}
Wood W~M, Siders C~W and Downer M~C 1991 {\em Phys. Rev. Lett.\/} {\bf 67}(25)
  3523

\bibitem{mendoncca2000book}
Mendon{\c{c}}a J~T 2000 {\em Theory of photon acceleration\/} (Institute of
  Physics Publishing, wholly owned by The Institute of Physics, London)

\bibitem{Mendonca2002}
Mendon{\c{c}}a J~T and Shukla P~K 2002 {\em Physica Scripta\/} {\bf 65} 160

\bibitem{Kenan_slow_ionization}
Qu K and Fisch N~J 2019 {\em Phys. Plasmas\/} {\bf 26} 083105

\bibitem{Shvets2017}
Shcherbakov M~R, Werner K, Fan Z, Talisa N, Chowdhury E and Shvets G 2019 {\em
  Nat. Commun.\/} {\bf 10} 1345

\bibitem{Nishida2012}
Nishida A, Yugami N, Higashiguchi T, Otsuka T, Suzuki F, Nakata M, Sentoku Y
  and Kodama R 2012 {\em Appl. Phys. Lett.\/} {\bf 101} 161118

\bibitem{Kenan_2018_upshift}
Qu K, Jia Q, Edwards M~R and Fisch N~J 2018 {\em Phys. Rev. E\/} {\bf 98}(2)
  023202

\bibitem{Edwards_Chirped}
Edwards M~R, Qu K, Jia Q, Mikhailova J~M and Fisch N~J 2018 {\em Phys.
  Plasmas\/} {\bf 25} 053102

\bibitem{Bulanov2005}
Bulanov S~S, Fedotov A~M and Pegoraro F 2005 {\em Phys. Rev. E\/} {\bf 71}(1)
  016404

\bibitem{Peng_PRApp2021}
Peng H, Riconda C, Weber S, Zhou C~T and Ruan S~C 2021 {\em Phys. Rev.
  Applied\/} {\bf 15}(5) 054053

\bibitem{jirka_electron_2016}
Jirka M, Klimo O, Bulanov S~V, Esirkepov T~Z, Gelfer E, Bulanov S~S, Weber S
  and Korn G 2016 {\em Phys. Rev. E\/} {\bf 93} 023207

\bibitem{grismayer_seeded_2017}
Grismayer T, Vranic M, Martins J~L, Fonseca R~A and Silva L~O 2017 {\em Phys.
  Rev. E\/} {\bf 95} 023210

\bibitem{Griffith2022}
Griffith A, Qu K and Fisch N~J 2022 {\em Physics of Plasmas\/} {\bf 29} 073104

\bibitem{PRL_LWFA2019}
Gonsalves A~J, Nakamura K, Daniels J, Benedetti C, Pieronek C, de~Raadt T~C~H,
  Steinke S, Bin J~H, Bulanov S~S, van Tilborg J, Geddes C~G~R, Schroeder C~B,
  T\'oth C, Esarey E, Swanson K, Fan-Chiang L, Bagdasarov G, Bobrova N, Gasilov
  V, Korn G, Sasorov P and Leemans W~P 2019 {\em Phys. Rev. Lett.\/} {\bf
  122}(8) 084801

\bibitem{Andreev2003}
Andreev N~E, Kuznetsov S~V, Pogosova A~A, Steinhauer L~C and Kimura W~D 2003
  {\em Phys. Rev. ST Accel. Beams\/} {\bf 6}(4) 041301

\bibitem{Welch2017}
Welch J~R, Zgadzaj R, Polyanskiy M, Zhang C, Pogorelsky I and Downer M~C 2017
  Mid-ir, co2-laser driven, self-modulated wakes {\em Conference on Lasers and
  Electro-Optics\/} (Optica Publishing Group) p FM2D.6

\bibitem{Kumar2021}
Kumar P, Yu K, Zgadzaj R, Downer M, Petrushina I, Samulyak R, Litvinenko V and
  Vafaei-Najafabadi N 2021 {\em Physics of Plasmas\/} {\bf 28} 013102

\bibitem{Brunetti2022}
Brunetti E, Campbell R~N, Lovell J and Jaroszynski D~A 2022 {\em Scientific
  Reports\/} {\bf 12} 6703 ISSN 2045-2322

\bibitem{felsen1970wave}
Felsen L and Whitman G 1970 {\em IEEE Transactions on Antennas and
  Propagation\/} {\bf 18} 242--253

\bibitem{kravtsov1974geometrical}
Kravtsov Y~A, Ostrovsky L~A and Stepanov N~S 1974 {\em Proceedings of the
  IEEE\/} {\bf 62} 1492--1510

\bibitem{AuYeung:83}
AuYeung J~C 1983 {\em Opt. Lett.\/} {\bf 8} 148--150

\bibitem{stepanov1993waves}
Stepanov N~S 1993 {\em Radiophysics and quantum electronics\/} {\bf 36}
  401--409

\bibitem{ostrovskii1971review}
Ostrovskii L~A and Stepanov N~S 1971 {\em Radiophysics and Quantum
  Electronics\/} {\bf 14} 387--419

\bibitem{jiang1975wave}
Jiang C~L 1975 {\em IEEE Transactions on Antennas and Propagation\/} {\bf 23}
  83--90

\bibitem{lampe1978interaction}
Lampe M, Ott E and Walker J~H 1978 {\em The Physics of Fluids\/} {\bf 21}
  42--54

\bibitem{wilks1989photon}
Wilks S~C, Dawson J~M, Mori W~B, Katsouleas T and Jones M~E 1989 {\em Phys.
  Rev. Lett.\/} {\bf 62} 2600

\bibitem{mori1991generation}
Mori W 1991 {\em Physical Review A\/} {\bf 44} 5118

\bibitem{kalluri1992frequency}
Kalluri D and Goteti V 1992 {\em Journal of applied physics\/} {\bf 72}
  4575--4580

\bibitem{savage1993frequency}
Savage R, Brogle R, Mori W and Joshi C 1993 {\em IEEE transactions on plasma
  science\/} {\bf 21} 5--19

\bibitem{yablonovitch1973spectral}
Yablonovitch E 1973 {\em Phys. Rev. Lett.\/} {\bf 31} 877

\bibitem{joshi1990demonstration}
Joshi C~J, Clayton C, Marsh K, Hopkins D, Sessler A and Whittum D 1990 {\em
  IEEE Transactions on Plasma Science\/} {\bf 18} 814--818

\bibitem{kuo1990frequency}
Kuo S~P 1990 {\em Phys. Rev. Lett.\/} {\bf 65} 1000

\bibitem{yugami2002experimental}
Yugami N, Niiyama T, Higashiguchi T, Gao H, Sasaki S, Ito H and Nishida Y 2002
  {\em Phys. Rev. E\/} {\bf 65} 036505

\bibitem{Suckewer2002}
Geltner I, Avitzour Y and Suckewer S 2002 {\em Appl. Phys. Lett.\/} {\bf 81}
  226

\bibitem{Suckewer2005}
Avitzour Y, Geltner I and Suckewer S 2005 {\em J. Phys. B\/} {\bf 38} 779

\bibitem{DEXY}
Abramowicz H, Altarelli M, Aßmann R, Behnke T, Benhammou Y, Borysov O,
  Borysova M, Brinkmann R, Burkart F, Büßer K, Davidi O, Decking W, Elkina N,
  Harsh H, Hartin A, Hartl I, Heinemann B, Heinzl T, TalHod N, Hoffmann M,
  Ilderton A, King B, Levy A, List J, Maier A~R, Negodin E, Perez G, Pomerantz
  I, Ringwald A, Rödel C, Saimpert M, Salgado F, Sarri G, Savoray I, Teter T,
  Wing M and Zepf M 2019 Letter of intent for the luxe experiment

\end{thebibliography}

\end{document}